\newcommand{\CLASSINPUTtoptextmargin}{19mm}%
\newcommand{\CLASSINPUTbottomtextmargin}{43mm}%
\newcommand{\CLASSINPUTinnersidemargin}{12.9mm}%
\newcommand{\CLASSINPUToutersidemargin}{12.9mm}%
\documentclass[conference,10pt,a4paper]{IEEEtran}
\usepackage{amsmath}
\usepackage{times}
\usepackage{graphicx}
\usepackage{multirow}
\usepackage[none]{hyphenat}
\usepackage{float}
\usepackage{subfig}
\usepackage{iftex}
\usepackage{xcolor}
\usepackage{siunitx}
\DeclareSIUnit{\decibel}{dB}
\DeclareSIUnit{\dBic}{dBic}
\usepackage{tcolorbox}
\tcbuselibrary{breakable}
\usepackage[backend=bibtex,defernumbers=true,sorting=none,style=ieee]{biblatex}%
\usepackage{t1enc}
\usepackage{times}

\makeatletter

\def\@maketitle{\newpage
\bgroup\par\addvspace{0.5\baselineskip}\centering%
\ifCLASSOPTIONtechnote
   {\bfseries\large\@IEEEcompsoconly{\sffamily}\@title\par}\vskip 1.3em{\lineskip .5em\@IEEEcompsoconly{\sffamily}\@author
   \@IEEEspecialpapernotice\par{\@IEEEcompsoconly{\vskip 1.5em\relax
   \@IEEEtitleabstractindextextbox{\@IEEEtitleabstractindextext}\par
   \hfill\@IEEEcompsocdiamondline\hfill\hbox{}\par}}}\relax
\else
   \vskip0.2em{\APMCtitlesize\ifCLASSOPTIONtransmag\bfseries\LARGE\fi\@IEEEcompsoconly{\sffamily}\@IEEEcompsocconfonly{\normalfont\normalsize\vskip 2\@IEEEnormalsizeunitybaselineskip
   \bfseries\Large}\@title\par}\vskip1.0em\par
   \ifCLASSOPTIONconference%
      {\@IEEEspecialpapernotice\mbox{}\vskip\@IEEEauthorblockconfadjspace%
       \mbox{}\hfill\begin{@IEEEauthorhalign}\@author\end{@IEEEauthorhalign}\hfill\mbox{}\par}\relax
   \else
      \ifCLASSOPTIONpeerreviewca
         {\@IEEEcompsoconly{\sffamily}\@IEEEspecialpapernotice\mbox{}\vskip\@IEEEauthorblockconfadjspace%
          \mbox{}\hfill\begin{@IEEEauthorhalign}\@author\end{@IEEEauthorhalign}\hfill\mbox{}\par
          {\@IEEEcompsoconly{\vskip 1.5em\relax
           \@IEEEtitleabstractindextextbox{\@IEEEtitleabstractindextext}\par\hfill
           \@IEEEcompsocdiamondline\hfill\hbox{}\par}}}\relax
      \else
         \ifCLASSOPTIONtransmag
           {\@IEEEspecialpapernotice\mbox{}\vskip\@IEEEauthorblockconfadjspace%
            \mbox{}\hfill\begin{@IEEEauthorhalign}\@author\end{@IEEEauthorhalign}\hfill\mbox{}\par
           {\vspace{0.5\baselineskip}\relax\@IEEEtitleabstractindextextbox{\@IEEEtitleabstractindextext}\vspace{-1\baselineskip}\par}}\relax
         \else
           {\lineskip.5em\@IEEEcompsoconly{\sffamily}\sublargesize\@author\@IEEEspecialpapernotice\par
           {\@IEEEcompsoconly{\vskip 1.5em\relax
            \@IEEEtitleabstractindextextbox{\@IEEEtitleabstractindextext}\par\hfill
            \@IEEEcompsocdiamondline\hfill\hbox{}\par}}}\relax
         \fi
      \fi
   \fi
\fi\par\addvspace{0.0\baselineskip}\egroup}

\def\APMCtitlesize{\@setfontsize{\APMCtitlesize}{24}{24pt}}
\def\APMCauthorsize{\@setfontsize{\APMCauthorsize}{11}{11pt}}
\def\APMCaffilsize{\@setfontsize{\APMCaffilsize}{10}{10pt}}
\def\APMCcaptionsize{\@setfontsize{\APMCcaptionsize}{9}{10pt}}
\def\APMCbibsize{\@setfontsize{\APMCbibsize}{8}{10pt}}

\def\@IEEEauthorblockNstyle{\APMCauthorsize\@IEEEcompsocnotconfonly{\sffamily}\@IEEEcompsocconfonly{\large}}
\def\@IEEEauthorblockAstyle{\APMCaffilsize\@IEEEcompsocnotconfonly{\sffamily}\@IEEEcompsocconfonly{\itshape}\@IEEEcompsocconfonly{\large}}
\def\@IEEEauthordefaulttextstyle{\APMCauthorsize\@IEEEcompsocnotconfonly{\sffamily}\sublargesize}

\def\thebibliography#1{\section*{\refname}%
    \addcontentsline{toc}{section}{\refname}%
    \APMCbibsize\@IEEEcompsocconfonly{\small}\vskip 0.3\baselineskip plus 0.1\baselineskip minus 0.1\baselineskip
    \list{\@biblabel{\@arabic\c@enumiv}}%
    {\settowidth\labelwidth{\@biblabel{#1}}%
    \leftmargin\labelwidth
    \advance\leftmargin\labelsep\relax
    \itemsep \IEEEbibitemsep\relax
    \usecounter{enumiv}%
    \let\p@enumiv\@empty
    \renewcommand\theenumiv{\@arabic\c@enumiv}}%
    \let\@IEEElatexbibitem\bibitem%
    \def\bibitem{\@IEEEbibitemprefix\@IEEElatexbibitem}%
\def\newblock{\hskip .11em plus .33em minus .07em}%
\ifCLASSOPTIONtechnote\sloppy\clubpenalty4000\widowpenalty4000\interlinepenalty100%
\else\sloppy\clubpenalty4000\widowpenalty4000\interlinepenalty500\fi%
    \sfcode`\.=1000\relax}

\long\def\@makecaption#1#2{%
\ifx\@captype\@IEEEtablestring%
\par\@IEEEtabletopskipstrut
\else
\@IEEEfigurecaptionsepspace
\fi
\setbox\@tempboxa\hbox{\normalfont\footnotesize {#1.}\nobreakspace\nobreakspace #2}%
\ifdim \wd\@tempboxa >\hsize%
\setbox\@tempboxa\hbox{\normalfont\footnotesize {#1.}\nobreakspace\nobreakspace}%
\parbox[t]{\hsize}{\normalfont\footnotesize\noindent\unhbox\@tempboxa#2}%
\else
\ifCLASSOPTIONconference \hbox to\hsize{\normalfont\footnotesize\hfil\box\@tempboxa\hfil}%
\else \hbox to\hsize{\normalfont\footnotesize\box\@tempboxa\hfil}%
\fi\fi
\ifx\@captype\@IEEEtablestring%
\@IEEEtablecaptionsepspace
\else
\fi}

\newlength\tablecaptiontotableskip
\newlength\figuretocaptionskip
\def\@IEEEfigurecaptionsepspace{\vskip\figuretocaptionskip\relax}%
\def\@IEEEtablecaptionsepspace{\vskip\tablecaptiontotableskip\relax}%

\def\abstract{\normalfont%
\@IEEEabskeysecsize\bfseries\textit{\abstractname}\,\bfseries\textit{---}\,%
\@IEEEgobbleleadPARNLSP}%

\def\IEEEkeywords{\normalfont%
\@IEEEabskeysecsize\bfseries\textit{\IEEEkeywordsname}\,\bfseries\textit{---}\,%
\@IEEEgobbleleadPARNLSP}%
\def\endIEEEkeywords{\relax\vspace{0.67ex}%
\par\if@twocolumn\else\endquotation\fi%
\normalsize\normalfont}%

\DeclareRobustCommand*{\APMCauthorrefmark}[1]{\raisebox{0pt}[0pt][0pt]{\textsuperscript{#1}}}%
\def\@IEEEauthorblockNtopspace{0ex}
\def\@IEEEauthorblockAtopspace{1mm}
\def\tablename{Table}
\def\thetable{\arabic{table}}
\def\IEEEkeywordsname{Keywords}
\def\subsubsection{\@startsection{subsubsection}{3}{\z@}{1.5ex plus 1.5ex minus 0.5ex}%
{0.7ex plus .5ex minus 0ex}{\normalfont\normalsize\itshape}}%
\newlength{\CPheadmatchindent}%
\def\@seccntformat#1{\hbox to\CPheadmatchindent{\csname the#1dis\endcsname}\hskip 0.1em \relax}
\IEEEilabelindentA \parindent
\IEEEilabelindent \IEEEilabelindentA
\IEEEelabelindent \parindent
\IEEEdlabelindent \parindent
\IEEElabelindent \parindent
\makeatother

\newcommand\copyrighttext{%
\footnotesize \textcopyright~2026 IEEE. Personal use of this material is permitted.  Permission from IEEE must be obtained for all other uses, in any current or future media, including reprinting/republishing this material for advertising or promotional purposes, creating new collective works, for resale or redistribution to servers or lists, or reuse of any copyrighted component of this work in other works.}

\newcommand\copyrightnotice{%
\begin{tikzpicture}[remember picture,overlay]
\node[anchor=south,yshift=10pt] at (current page.south) {\fbox{\parbox{\dimexpr0.75\textwidth-\fboxsep-\fboxrule\relax}{\copyrighttext}}};
\end{tikzpicture}%
}

\begin{document}
\raggedbottom
%
%
%
\title{From Prompt to Prototype: Towards a Frontier LLM Driven RF Engineering Workflow}
%
%
\author{%
\IEEEauthorblockN{%
M. Heinrichs\APMCauthorrefmark{\#},
O. Moschner\APMCauthorrefmark{\$},
S. Tewes\APMCauthorrefmark{\#},
V. Wienstroer\APMCauthorrefmark{\$},
A. Sezgin\APMCauthorrefmark{\#},
R. Kronberger\APMCauthorrefmark{\$}
}
\IEEEauthorblockA{%
\APMCauthorrefmark{\#}Digital Communication Systems, Ruhr-University Bochum, Germany\\
\APMCauthorrefmark{\$}RF-Laboratory, TH Köln - University of Applied Sciences, Germany\\
markus.heinrichs@rub.de
}
}
%
\maketitle
%
%
%
\begin{abstract}
Agentic coding environments give a frontier large language model (LLM) direct access to a workstation's terminal, file system, and software. This work demonstrates they extend to professional RF hardware design: an active GNSS L1-band antenna -- a circularly polarized patch, surface acoustic wave (SAW) prefilter, and two-stage low-noise amplifier (LNA) on one printed circuit board (PCB) -- was designed, optimized, and made manufacturing-ready. The LLM agent autonomously operated CST Studio Suite, Keysight ADS, and KiCad via scripting interfaces. Engineer input was limited to the specification, trade-off decisions, and design reviews. Workflow, results, and the RF engineer's evolving role are discussed.
\copyrightnotice
\end{abstract}
%
%

\section{Introduction}
Since their public release as chat assistants, large language models (LLMs) have evolved into \emph{agentic} AI systems: models that autonomously plan, execute, and verify multi-step tasks by utilizing their ability to operate external tools \cite{acharya2025agentic}. The most widely used implementation of this approach is the agentic coding environment, such as Claude Code \cite{anthropic_claudecode}, which grants a frontier LLM direct access to the terminal, the file system, and the programs installed on a local workstation. These environments were built for software engineering, where the agent writes code, executes it, inspects the result, and iterates until a specified goal is reached.

In contrast, hardware design has previously been approached using domain-specific AI: reinforcement-learning and inverse-design methods synthesize RF integrated circuits \cite{sengupta2026spectrum}, LLM-based agents orchestrate digital design flows spanning synthesis, placement, and routing down to tape-out \cite{wu2024chateda}, and electronic design automation (EDA) vendors integrate proprietary AI assistants into their own suites. All of these require custom-built tools or vendor support. Professional RF design software, however, already exposes exactly the interfaces that a coding agent operates natively, such as documented file formats, command-line executables, and Python APIs. RF design can therefore be reformulated as a coding task: the agent scripts the simulators instead of interacting with their graphical user interface, and the complete design state becomes regenerable source code without any AI integration on the tool vendor's side.

\begin{figure}[htb]
\centering
\includegraphics[width=0.9\linewidth]{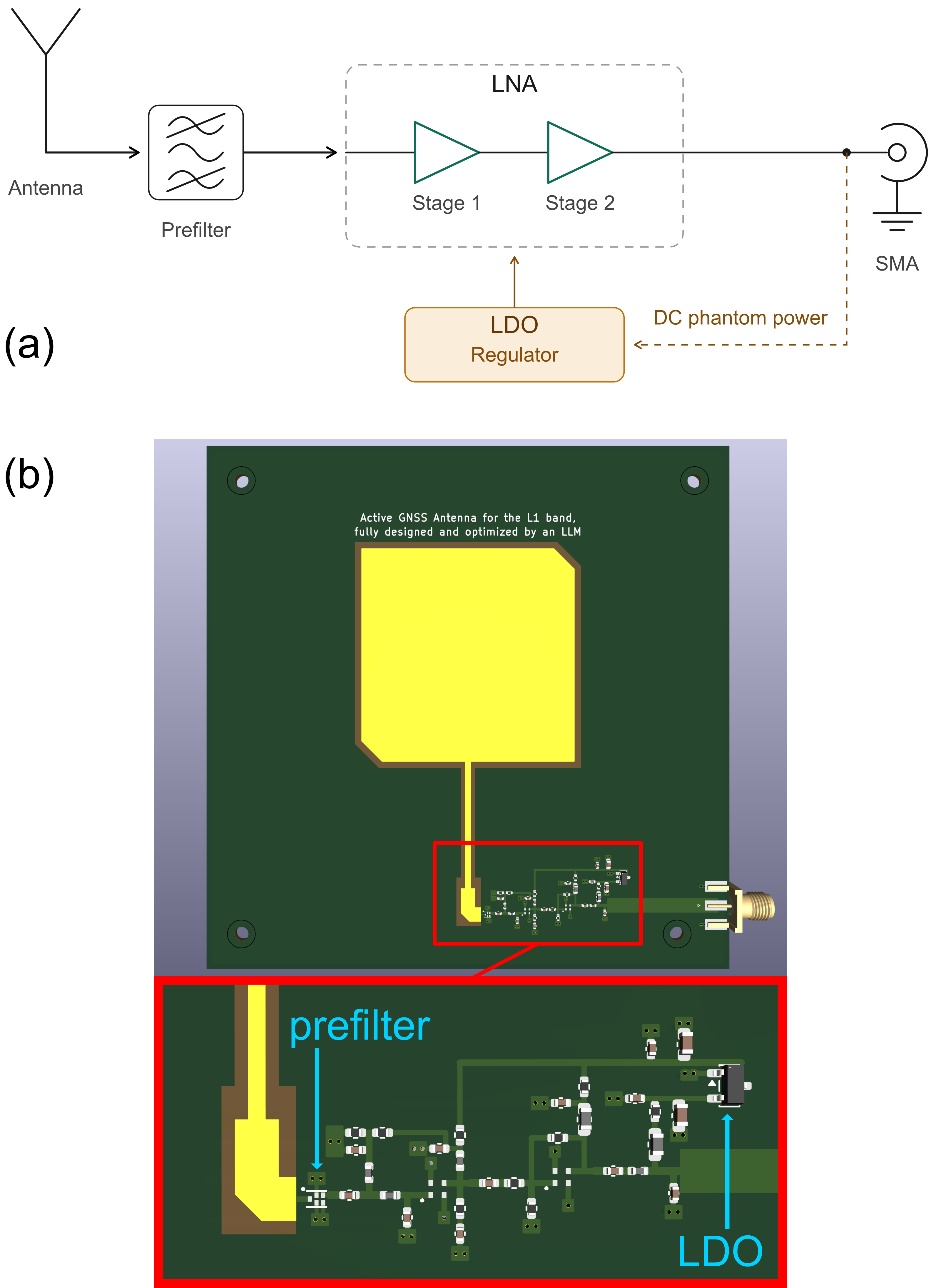}\\
\caption{(a) Block diagram; (b) KiCad rendering of the final active GNSS antenna as generated by the LLM agent, with detail view of the circuit section.}
\label{fig:pcb}
\vspace{-\baselineskip}
\end{figure}

This paper demonstrates that this new approach is already working in practice today, not just for a specific subtask, but for a complete product design. Starting from a single specification prompt, the frontier model Claude Fable 5 \cite{anthropic2026fable} operating inside Claude Code designed an active GNSS antenna for the L1 band as shown in the block diagram in Fig.~\ref{fig:pcb}a. The AI agent did all the necessary steps: it selected the substrate and performed the full-wave design of a circularly polarized patch antenna in CST, designed a two-stage LNA with SAW prefilter in ADS, draw the PCB layout in KiCad and finally generated a complete fabrication data set resulting in a manufacturing-ready board shown in Fig.~\ref{fig:pcb}b.

\section{Agentic Design Workflow} \label{sec:workflow}
The design was carried out on a standard engineering workstation running CST Studio Suite 2026, ADS 2024, and KiCad 9. The agent Claude Fable 5 running inside the Claude Code command-line environment operates these tools exclusively through their scripting interfaces, specifically the CST Python libraries, ADS netlists executed by \texttt{hpeesofsim}, and KiCad's \texttt{pcbnew} module and \texttt{kicad-cli}. With a runtime of \SI{2}{}--\SI{3}{\second} per circuit simulation and \SI{3}{}--\SI{5}{\minute} per fine-mesh full-wave simulation, the agent can afford hundreds of simulate--evaluate--modify iterations.

A new LLM session starts without any memory of previous ones, so an agent would ordinarily have to familiarize itself with the automation of CST and ADS at the beginning of every single project. To bypass this repetitive overhead, context input is utilized, providing essential operating knowledge upfront -- in this case as two short markdown documents summarizing the Python interfaces of the two simulators. These context files originate from previous sessions, where they were created and iteratively maintained by the corresponding agent. All further knowledge, from design methodology to component selection, originates from the model itself.

The prompt, shown in Fig.~\ref{fig:prompt}, reflects a discipline that is well known to every RF engineer, where the specification is fixed before development starts. It sets the boundary conditions, which are a single two-layer PCB with a maximum outline of $\SI{120}{\milli\meter} \times \SI{120}{\milli\meter}$, phantom powering with \SI{5}{}--\SI{12}{\volt} over the RF port and components available at the mandated distributors in this case, followed by the functional requirements for antenna and LNA discussed in Section~\ref{sec:design}.

The completion criteria at the end of the prompt are equally familiar: all specifications met in simulation, a KiCad project with clean electrical rule check (ERC) and design rule check (DRC) consistent with the simulated design, a visual inspection of the final layout, and complete fabrication data. These are the same acceptance gates a human engineer works toward, while for the agent, they additionally define when its autonomous loop may terminate.

\begin{figure}[htb]
\begin{tcolorbox}[colback=gray!8, colframe=gray!50, title=Initial Prompt, fonttitle=\bfseries, left=2mm, right=2mm, top=1mm, bottom=1mm]
\ttfamily\scriptsize
Your task is to design and optimize an active GNSS antenna for the L1 band on a single double-layerd Printed Circuit Board (PCB) that is ready for manufacturing and assembly at JLCPCB.\\
General specification:
\begin{itemize}
    \item Single PCB solution with two copper layers and maximum outline of 120 x 120 mm.
    \item The PCB contains the antenna, LNA circuitry and an SMA connector.
    \item Phantom power is supplied to the RF port (5 to 12 Volt).
    \item It is up to you to choose from the available substrate material and thickness.
    \item All components have to be available at both distributors Mouser and LCSC, so that JLC can deliver a fully assembled antenna.
    \item Place the following text on the silkscreen: ``Active GNSS Antenna for the L1 band, fully designed and optimized by an LLM''
\end{itemize}
LNA specification:\\
...\\
Antenna specification:\\
...\\
Your task is done, when all of the following criteria are met:\\
...\\
If you have any questions, try to clarify them at the beginning rather than in the middle of the development process.\\
For context on how to use CST and ADS, read the following markdown files first:
\begin{itemize}
    \item ADS\_Python\_Interface\_Reference.md
    \item CST\_Python\_Interface\_Reference.md
\end{itemize}
\end{tcolorbox}
\caption{Initial prompt, the detailed LNA and antenna specifications and the completion criteria are abbreviated here.}
\label{fig:prompt}
\end{figure}

In contrast to a common human design process, the agent maintains the entire design state as scripts and parameter files, so that schematic, board, design-rule checks, renders, and fabrication data regenerate with a single command.

\section{Design Case: Active GNSS L1 Antenna} \label{sec:design}
The final design from Fig.~\ref{fig:pcb} integrates a right-hand circularly polarized (RHCP) corner-truncated patch antenna, a quarter-wave transformer, a SAW prefilter, a two-stage LNA, and an edge-mounted SMA connector. The specification and all simulated results are listed in Table~\ref{tab:results}. Rather than documenting the design in full detail, the following subsections use the individual design steps to show how the agent reasons, decides, and verifies.

\subsection{Substrate Selection and Antenna}
Throughout the project, the agent faced the same design decisions as a human engineer, except for where the engineer draws on experience, the agent has to generate its own evidence. The substrate choice, deliberately left open in the prompt, illustrates this. An experienced RF designer selects a low-loss RF laminate for a GNSS patch antenna as a matter of course. The agent arrived at the same choice by producing data: it first modeled the patch on standard FR4 substrate, where its simulation showed the dielectric loss limiting the radiation efficiency to \SI{28}{\percent}, which is too little for the required \SI{3}{\dBic} of realized gain, and then repeated the analysis on Rogers RO4350B from the manufacturer's high-frequency portfolio, which lifted the efficiency to \SI{78}{\percent}. Only then did it commit to the RF laminate.

\begin{figure}[htb]
\centering
\includegraphics[width=0.74\linewidth]{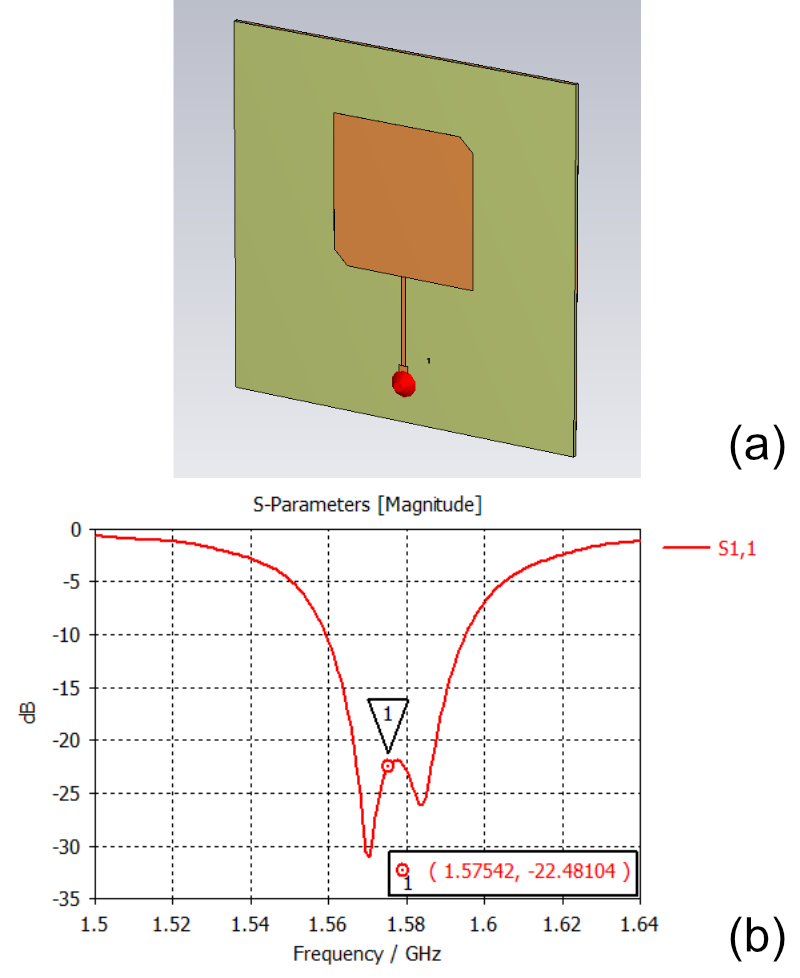}
\caption{CST simulation: (a) 3D model of the final antenna; (b) reflection coefficient at the antenna feed.}
\label{fig:antenna}
\vspace{-\baselineskip}
\end{figure}

For circular polarization, the agent selected a corner-truncated square patch \cite{sharma1983cp}, the standard single-feed solution. Its optimization provides the most insightful finding: the axial ratio initially appeared to be stuck at \SI{3}{\decibel}, exactly at the specification limit. Instead of accepting the marginal result, the agent challenged his own methodology. An auxiliary probe-fed model ruled out the microstrip feed as the cause, while the actual problem turned out to be the agent's sequential one-parameter-at-a-time search combined with a too coarse frequency sampling. A joint two-dimensional optimization of patch length and corner truncation at the exact center frequency resolved the issue. This makes the agent's debugging behavior visible when applied to an electromagnetic problem -- hypothesis, isolation experiment, correction. The final antenna achieves an axial ratio of \SI{2.46}{\decibel} and \SI{5.15}{\dBic} realized gain at boresight. The 3D model from CST as well as the simulated return loss of the antenna are shown in Fig.~\ref{fig:antenna}.

\subsection{Low-Noise Amplifier}
For the LNA, the specification prescribed the transistor model that has to be used, an Infineon BFP840FESD \cite{infineon_bfp840fesd}. Mandating a component like this is a deliberate instrument of the specification, as it reduces the number of possible solutions for the agent. Here, the engineer prescribed a transistor known from prior experience, but availability, qualification, or company standards can equally motivate such human design choices.

Two observations from this phase are worth relating. The first shows the agent's diligence with component documentation: while reviewing the transistor datasheet mid-design, it noticed that the ESD-protected variant is limited to $V_\mathrm{CEO} = \SI{2.25}{\volt}$, a limit its initially chosen bias point violated, and redesigned the biasing accordingly, without being asked to.

\begin{figure}[htb]
\centering
\includegraphics[width=0.65\linewidth]{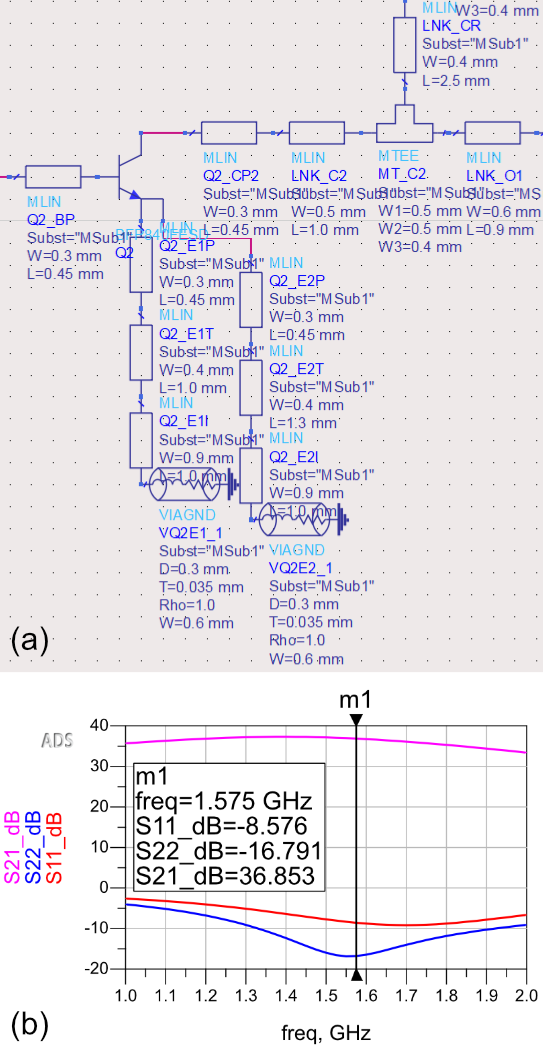}
\caption{ADS simulation: (a) schematic (detail view); (b) simulated S-parameters of the final LNA.}
\label{fig:ads}
\vspace{-\baselineskip}
\end{figure}

The second observation concerns the schematic itself and the value of human review. The agent's first schematic was functionally complete but not human-readable, since every component sat on a regular grid, with connectivity expressed solely through net labels at the pin ends. This is adequate for a machine but of little use for a human design review. When the engineer requested a conventional signal-flow arrangement with drawn wires, the agent rebuilt the schematic geometrically -- cosmetic imperfections such as obscured labels remain -- and, rather than trusting its own artifacts, compared the old and new netlists as graphs. Feedback of this kind, once given, does not need to be repeated, as agentic coding environments support persistent context files, and a single instruction such as ``arrange schematics in signal-flow order with drawn wires'' turns the correction into a standing convention for all future projects. A detail view of the rearranged ADS schematic as well as scattering parameters of the final LNA are given in Fig.~\ref{fig:ads}.

\subsection{SAW Prefilter and System Performance}
The prefilter demonstrates how the agent absorbs changing requirements. When the engineer requested an L1 prefilter mid-project, the agent processed it like a design change request. It shortlisted GPS SAW filters available at both mandated distributors and again decided on generated data, selecting the TDK B4300 \cite{tdk_b4300} over a lower-loss alternative because \SI{0.15}{\decibel} of additional insertion loss result in \SI{4}{}--\SI{7}{\decibel} more rejection exactly in the threatened cellular and WiFi bands. Subsequently, the agent confirmed that the LNA still complies with all specifications, as summarized alongside the simulated results in Table~\ref{tab:results}.

\begin{table}[htb]
\renewcommand{\arraystretch}{1.2}
\caption{Specification and simulated performance.}
\label{tab:results}
\centering
\begin{tabular}{|l|c|c|}
\hline
Parameter & Spec. & Simulated \\
\hline\hline
\multicolumn{3}{|l|}{\textit{Antenna (CST, full board, fine mesh)}} \\
\hline
Realized boresight gain & $\geq 3$\,dBic & 5.15\,dBic \\
Axial ratio at $f_0$ & $\leq 3$\,dB & 2.46\,dB \\
Reflection coefficient $|S_{11}|$ & --- & $-22.6$\,dB \\
\hline\hline
\multicolumn{3}{|l|}{\textit{LNA (ADS, as-built layout parasitics)}} \\
\hline
Gain at $f_0$ & $\geq 25$\,dB & 36.9\,dB \\
Noise figure & $< 1$\,dB & 0.77\,dB \\
Stability $k$ (\SI{1}{\mega\hertz}--\SI{12}{\giga\hertz}) & $> 1$ & $\geq 1.96$ \\
Output return loss $|S_{22}|$ & $\leq -10$\,dB & $-16.8$\,dB \\
\hline\hline
\multicolumn{3}{|l|}{\textit{System with B4300 prefilter}} \\
\hline
Gain / noise figure & --- & 35.5 / 1.69\,dB \\
$|S_{11}|$ / $|S_{22}|$ & --- & $-14.4$ / $-17.4$\,dB \\
Stability $k$ (\SI{1}{\mega\hertz}--\SI{12}{\giga\hertz}) & $>1$ & $\geq 4.1$ \\
Out-of-band improvement & --- & 41--51\,dB \\
\hline
\end{tabular}
\end{table}

\subsection{PCB Integration and Fabrication Data} \label{subsec:pcb}
\begin{figure}[htb]
\centering
\includegraphics[width=0.9\linewidth]{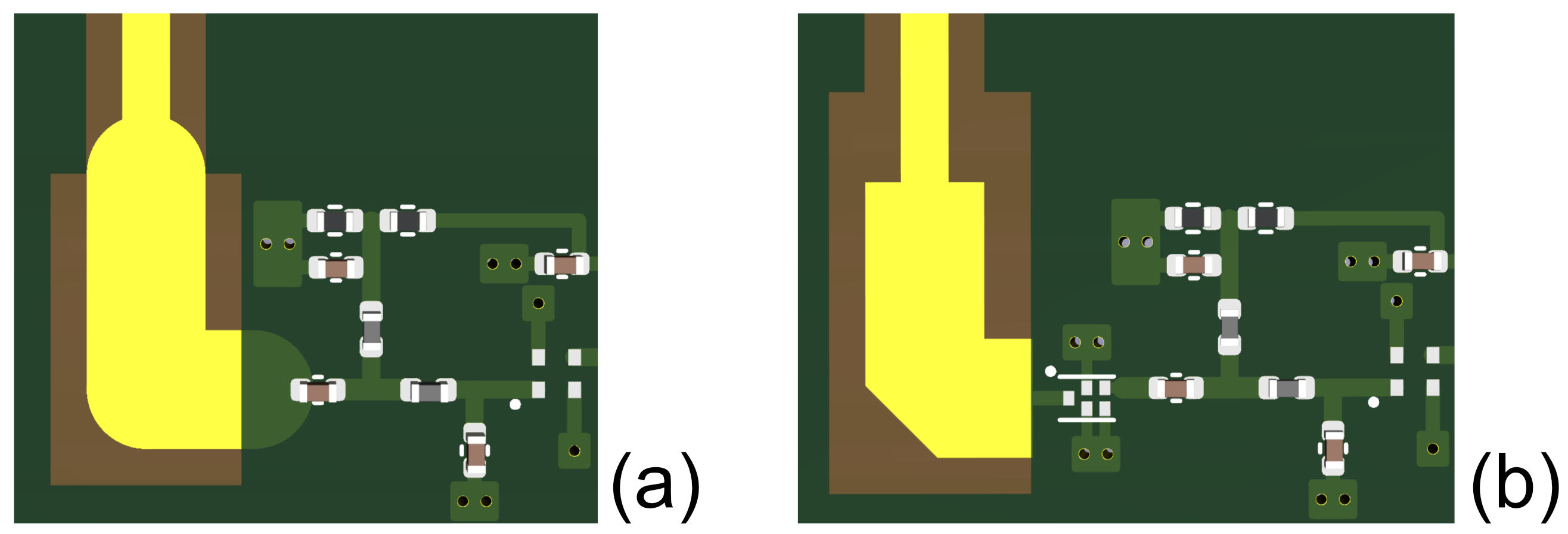}
\caption{KiCad design iterations: (a) first autonomous output; (b) final design after review-driven refinement.}
\label{fig:iterations}
\vspace{-\baselineskip}
\end{figure}

Because the entire KiCad project is generated from the scripted design description (Section~\ref{sec:workflow}), schematic and layout are netlist-identical by construction, and an automatic check verifies all 75~nodes and 18~nets against the simulated design of record. The same rigor connects layout and simulation: the RF lines are drawn as copper polygons that replicate the simulated line models, so the board that gets manufactured is the board that was verified. The remaining deviations were caught by the second verification instance, the human: the engineer's visual reviews flagged idealizations that the agent then corrected, such as the second amplifier stage collector junction, for example, first entered the simulation as a crossing but was routed as a tee and is now consistent in simulation as the microstrip tee \texttt{MT\_C2} visible in the schematic detail of Fig.~\ref{fig:ads}a. To give an idea on how the PCB layout evolved, Fig.~\ref{fig:iterations} contrasts the first autonomous layout with the final board and makes the human-requested changes directly visible: the line endings, initially rounded, were squared to match the rectangular simulation models, the \SI{90}{\degree} corner in the feed line received a \SI{45}{\degree} miter, the solder-mask opening around the RF lines was extended, and the final board carries the SAW prefilter that entered the design as the mid-project change request. The completion criteria from the prompt were then met to the letter: ERC and DRC pass without violations under the manufacturer's rule set, every component is in stock at both mandated distributors, and the delivered package contains all manufacturing data in the assembly service's format.

\section{Impact on the RF Engineering Workflow} \label{sec:impact}
At no point during the project did the engineer operate CST, ADS, or KiCad directly, nor was any familiarity with their user interfaces or scripting APIs required. Work that is classically distributed over several specialist roles -- antenna design, circuit design, layout, and procurement -- was executed by one agent in approx. two hours of time in one continuous context, which kept the domains mutually consistent.

The tasks the agent solved autonomously go well beyond routine automation. These are engineering steps that require judgment, executed on professional tools rather than simplified surrogates. The engineer's role concentrates on what remains genuinely human in the workflow: writing precise and testable specifications, arbitrating trade-offs, and verification. Expert oversight remained load-bearing, as the as-built fidelity corrections in Section~\ref{subsec:pcb} originated from human design reviews, but it shifts from continuously operating the tools to review gates backed by machine-checkable criteria. As a side effect, the design exists entirely as version-controlled code and regenerates deterministically -- reproducible and transferable in a way most interactive CAD sessions are not.

However, some limitations frame these results. All performance figures are simulation results, as the prototype is pending fabrication and measurement. Also, the approach presumes documented scripting interfaces and an engineer able to judge the plausibility of intermediate results, and in the authors' experience, only the generation of frontier models released since early to mid-2026 sustains such multi-hour tool-driving sessions reliably. Additionally, this capability is not exclusive to the used Claude agent: other frontier models of the trillion-parameter class, such as OpenAI's GPT Sol or Moonshot AI's KIMI K3, ship with comparable agentic coding environments and should be able to drive the same workflow.

\section{Conclusion} \label{sec:conclusion}
A complete active GNSS L1 antenna, consisting of a circularly polarized patch, a SAW prefilter, a two-stage LNA, and full fabrication data, was designed end-to-end in just two hours by a frontier LLM operating professional RF tools from within a general-purpose agentic coding environment. All specifications are met in simulation. More broadly, agentic coding environments are not restricted to software development: wherever engineering tools expose scripting interfaces, hardware design becomes an automatable, reviewable, and reproducible process without any vendor-side AI integration. The engineer does not disappear from this loop but moves up in it from operating the tools to specifying, deciding, and verifying through measurements, a task inherently beyond the scope of an LLM agent.


\printbibliography

\end{document}